\documentstyle[epsfig]{l-aa}
\input epsf

\def \rsun {\ifmmode$R$_{\odot}\else R$_{\odot}$\fi}

\def \hcm {\hbox {\ifmmode $ H atoms cm$^{-2}\else H atoms cm$^{-2}$\fi}}
\def\approxgt{\mathrel{\hbox{\rlap{\lower.55ex \hbox {$\sim$}}
        \kern-.3em \raise.4ex \hbox{$>$}}}}
\def\approxlt{\mathrel{\hbox{\rlap{\lower.55ex \hbox {$\sim$}}
        \kern-.3em \raise.4ex \hbox{$<$}}}}

\newcommand {\sax} {{\it BeppoSAX }}

\newcommand {\etal} {et~al. }

\newcommand {\ergcms} {erg cm$^{-2}$ s$^{-1}~$}
\newcommand {\chisq} {$\chi ^{2}$}
\newcommand {\rchisq} {$\chi_{\nu} ^{2}$}

\def\lsim{\lower.5ex\hbox{$\; \buildrel < \over \sim \;$}}
\def\gsim{\lower.5ex\hbox{$\; \buildrel > \over \sim \;$}}

\begin{document}

   \thesaurus{         
              (; 
               )} 
   \title{Synchrotron and Inverse Compton Variability in the BL Lacertae 
   object S5~0716+714}

%   \subtitle{}

   \author{P. Giommi\inst{1}, E. Massaro\inst{2}, L. Chiappetti\inst{3}, 
 E.C. Ferrara\inst{4}, G., Ghisellini\inst{5}, Minhwan Jang\inst{6},
 M. Maesano\inst{2}, H.R. Miller\inst{4}, F. Montagni\inst{2},
 R. Nesci\inst{2}, P. Padovani\inst{7,10,11}, E. Perlman\inst{7},
 C.M. Raiteri\inst{8}, S. Sclavi\inst{2}, G. Tagliaferri\inst{5},
 G. Tosti\inst{9}, M. Villata\inst{8}}

   \offprints{P. Giommi}

\institute{
   {\inst{1}\sax Science Data Center, ASI,
    Via Corcolle, 19,
    I-00131 Roma , Italy}\\
   {\inst{2}Istituto Astronomico,
    Universita' "La Sapienza", Via Lancisi 29 
    I-00161 Roma, Italy}\\
   {\inst{3}IFC, CNR, 
     Via Bassini 15, 
     Milan, Italy}\\
    {\inst{4}Department of Physics and Astronomy,
     Georgia State University
     Atlanta, GA 30303, USA}\\
   {\inst{5}Osservatorio Astronomico di Brera,
    Via Bianchi 46,
    I-23807 Merate, Italy}\\
    {\inst{6}Dept. Astronomy and Space Science,
     Kyung Hee University,
     Yongin, Kyungki-Do 449-701, Korea}\\
   {\inst{7}Space Telescope Science Institute
   3700 San Martin Drive,
    Baltimore,  MD 21218
    U.S.A}\\
   {\inst{8}Osservatorio Astronomico di Torino,
     Strada Osservatorio 20
     I-10025 Pino Torinese,
     Italy} \\
   {\inst{9}Osservatorio Astronomico, 
     Universita' di Perugia,
     Perugia, Italy}\\
   {\inst{10}Affiliated to the Astrophysics Division, 
     Space Science Department,
     European Space Agency} \\
   {\inst{11}On leave of absence from Dipartimento di Fisica, 
     II Universit\`a di Roma ``Tor Vergata'', 
     Italy}\\
}
   \date{Received  ; accepted }

   \maketitle
   \markboth{Giommi, Massaro, et al., }{}

\begin{abstract}

We report intensity variations of different spectral components in 
the BL Lac object S5~0716+714 detected during coordinated 
\sax and optical observations in 1996 and 1998.
The transition between synchrotron and inverse Compton emission  
has been clearly detected as sharp X-ray spectral breaks at around 2-3
keV on both occasions. 
Correlated optical and soft X-ray variability was found during 
the second \sax pointing when intensive optical monitoring 
could be arranged.
The hard (Compton) component changed by a factor of 2 between 
the two observations, but remained stable within each exposure.  
During events of rapid variability S5~0716+714 showed 
spectral steepening with intensity, a behaviour rarely observed 
in BL Lacs. We interpret these findings as the probable consequence 
of a shift of the synchrotron peak emission from the IR/optical band 
to higher energies, causing the synchrotron tail to push into
the soft X-ray band more and more as the source brightens. 

    \keywords{galaxies, Active}

   \end{abstract}

\section{Introduction}

It is generally agreed that the main mechanism powering BL Lacs is
synchrotron emission followed by inverse Compton scattering in a
relativistic beaming scenario (e.g. Kollgaard 1994, Urry \& Padovani 1995,
Ghisellini, Maraschi \& Dondi  1996).
The synchrotron component peaks (in a $\nu~vs~\nu~f_\nu$ representation)
at energies ranging from infrared frequencies, for Low energy peaked
(LBL) BL Lacs (mostly discovered in radio surveys), to hard X-rays for 
extreme High energy peaked (HBL) BL Lacs (typically discovered 
in X-ray surveys, Padovani \& Giommi 1995, Sambruna, Maraschi \& Urry  1996). 

S5~0716+714 is a strongly variable (e.g. Wagner \& Witzel 1995, 
Otterbein \etal 1998) BL Lac object 
%discovered in a sample of bright flat spectrum radio 
%sources (K\"uhr \etal 1981) 
characterized by a spectral energy distribution (SED) peaking at 
intermediate frequencies ($\nu_{peak} \sim 10^{14}-10^{15}$ Hz) compared 
to LBL and HBL BL Lacs. 
To date no redshift has been measured for this object.

In this paper we report the detection of intensity and 
spectral variations involving both the synchrotron and the inverse 
Compton components of S5~0716+714.
The data are from two observations carried out with the Narrow 
Fields Instruments (NFI) of the \sax satellite (Boella et al. 1997a),
and from simultaneous optical observations made with a number of 
small telescopes in Italy and in Korea. 

\begin{table*}
\centerline{Table 1. - Log of the \sax observations of S5~0716+714 and 
image analysis results}
\begin{tabular}{lcccccc}
\hline \hline
Date    & LECS & Count rate & Count rate &  MECS & Count rate (2-10 keV) \\
        & exposure (s) & (0.1-2.keV)$ct~s^{-1}$ & (2-10 keV) $ct~s^{-1}$& exposure (s) & $ct~s^{-1}$ \\
\hline
14-NOV-1996 &  13122&$0.020\pm 0.001$ &$0.006\pm 0.0008$ & 122509& $0.0263^a\pm 0.0006$ \\
~~7-NOV-1998 &    ~9475&$0.022\pm 0.002$ &$0.011\pm 0.0010$ & ~~31317& $0.0344^b\pm
0.0012$ \\
\hline
\end{tabular}

$^a$ Three MECS units;~~$^b$ Two MECS units \\
\end{table*}

\section{BeppoSAX observations and data analysis}

S5~0716+714 was observed by \sax twice on November 14, 1996 and on  
November 7, 1998. 
Preliminary results on the first \sax observation have been reported in 
Chiappetti \etal (1997).

Screened event lists for the LECS (Parmar \etal 1997) and MECS (Boella et
al 1997b) instruments, and the average PDS (Frontera \etal 1997)
background-subtracted spectra were taken from the \sax SDC on-line archive
(Giommi \& Fiore 1998).
The data analysis was performed using the software available in the 
XANADU package (XIMAGE, XRONOS, XSPEC).

The analysis of the \sax X-ray images shows that S5~0716+714 was found 
in a rather faint state compared to previous observations (e.g. Urry \etal 
1996, Wagner \etal 1996). 
The count rates in the LECS and MECS instruments have been estimated 
using XIMAGE (Giommi et al. 1991), upgraded at 
the \sax SDC to support \sax imaging data. The observational parameters
and the measured count rates for the two instruments are given 
in Table 1.
Significant intensity variations were detected between and during each 
observation. 
In particular, the comparison of the 1996 and 1998 data shows that 
while the count rate in the soft band
(0.1-2 keV) was found at the same level, the intensity in the
harder 2-10 keV band changed by nearly a factor two both in the LECS
and in the MECS \footnote{The MECS instrument was operated 
with three units in 1996 and with two units in 1998; the 1998 count rate 
therefore must be multiplied by roughly 1.5 before being compared to 
the 1996 data} instruments.
This difference in variability implies that the X-ray spectral shape of 
S5~0716+714 changed significantly between the two observations.
On time scales of hours X-ray variability was detected only in the low 
energy band (0.1-2 keV) as discussed in section 4. 

Although some signal is present in the PDS data in both observations 
the source was too faint to be detected above the 
confusion limit of $\approx 2-3 \times 10^{-12}$ cgs. 

XSPEC spectral fits (over the full 0.1-10. keV LECS and MECS bands) with 
a single power law model, modified by Galactic 
absorption as estimated from 21 cm measurements ($N_H = 3.8\times 10^{20}$ 
cm$^{-2}$, Dickey and Lockman 1990) gave best fit energy spectral indices 
$\alpha_x(1996)=1.1\pm0.08$ and $\alpha_x(1998)=0.9\pm 0.1$
and with high \rchisq values (\rchisq = 1.35, 55 d.o.f. and \rchisq = 1.63, 
53 d.o.f, respectively) due to a poor fit at soft energies. 
These slopes are much flatter than the one found in a previous ROSAT PSPC 
observation ($\alpha_x=1.8$, 0.1-2.0 keV) during which S5~0716+714 was 
also highly variable (Urry \etal 1996, Wagner \etal 1996). 
We next fitted the data with a broken power law model and $N_H$ fixed 
to the Galactic value. 
The spectral fit and the residuals for the 1996 observation are shown 
in Fig. 1. The resulting spectral parameters with the statistical ($1\sigma$)
uncertainties are reported in Table 2.
In both observations the spectral slope is much steeper 
($\Delta \alpha = 0.6-0.7$ ) at lower energies (possibly consistent with the 
ROSAT slope), and the spectral break is around 2-3 keV.  
An F test shows that the broken power law model significantly improves 
(prob $> 99.99$\%) the fit compared to the single power law model. 
The \rchisq $~$ in the second observation ($\sim 1.5$) is still 
relatively high, likely because of the presence of non negligible 
rapid spectral variability (see below).  

\begin{table*}
\centering
\centerline{Table 2. - Results of LECS+MECS spectral analysis - Broken power 
law model}
\begin{tabular}{lcccccc}
\hline \hline
Date & $\alpha_{soft}$ & $\alpha_{hard}$& Break energy (keV)& \chisq (d.o.f.) 
& Flux 0.1-2 kev &Flux 2-10 kev\\
     & (energy index) & (energy index)   &  &   &  \ergcms & \ergcms \\
\hline
14-NOV-1996 & $1.7 \pm 0.3$ & $0.96 \pm 0.15$& $2.3\pm0.4$& 56(48) 
& $2.0\times 10^{-12} $&$1.4\times 10^{-12} $ \\
~~7-NOV-1998 & $1.3 \pm 0.4$ & $0.73 \pm 0.18$& $2.8\pm0.8$& 76(49)
& $1.8\times 10^{-12} $&$2.6\times 10^{-12} $ \\
\hline
\end{tabular}
\end{table*}

\begin{figure}
\epsfig{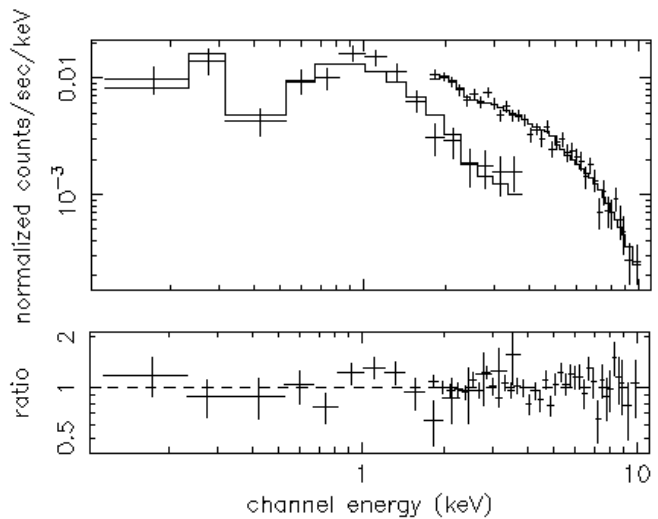}
\caption{LECS + MECS 1996 spectrum of S5~0716+714 fitted to a broken power law
and $N_H$ fixed to the Galactic value.}
\label{fig1}
\end{figure}

\begin{figure}
\epsfig{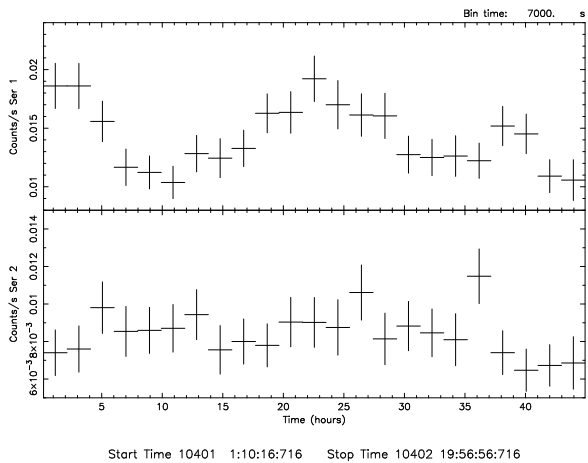}
\caption{The MECS lightcurve of S5~0716+714 in the 1.5-3.0 (upper panel) 
and 5.0-10 keV (lower panel)
energy bands during the 1996 \sax observation. Significant 
variability is present only at low energies.}
\label{fig2}
\end{figure}

\section{Optical observations}

Photometric measurements of S5~0716+714 were performed with telescopes
operated by the Roma, Perugia and Torino groups equipped with CCD cameras, 
two of them mounting a back illuminated SITe SIA502A chip.
The bandpasses used were the standard Johnson B, V, the Cousins R$_C$,
I$_C$ and the F86, F98 of the Arizona system (Johnson \& Mitchell 1975). 
BVR$_C$I$_C$ magnitudes of three standard stars in the field of S5~0716+714 
were taken from Ghisellini et al. (1997) and Villata \etal (1998a), while 
the magnitude in the 
Arizona filters of the same stars were calibrated with the primary standard 
BS2527 (Johnson \& Mitchell 1975). 
Other observations in the R$_C$ band were performed with the Kyung Hee 
University telescope in Korea at the beginning of the \sax pointing.
 
During both \sax X-ray observations the source was not in bright states 
and at about the same magnitude R$_C\sim$13.8.
The optical monitoring (Villata et al. 1998b, Massaro \etal 1999a, 
Raiteri \etal 1999) shows that in November 1996 S5~0716+714 was in a 
declining phase following a small burst occurred 
two weeks before, whereas in November 1998 it was in a mild brightening phase
twenty days after the lowest recorded level (R$_C$=14.40) since January 1996.

\section{Short-term variability}
Figure 2 shows the soft (1.3-3.0 keV) and hard (5.0-10. keV) MECS 
lightcurve during the long 1996 \sax observation. Clear variations of up to 
a factor of two are evident in the low energy curve, while above 
5 keV the source flux remains constant within approximately 20\%.
During this observation we were able to perform simultaneous 
optical monitoring (in the F86 Arizona band) for about five hours
starting at the beginning of the pointing, and covering only the 
initial part of the X-ray lightcurve. We observed a smooth decline of the 
luminosity: the F86 magnitude varied from 13.14 at 0.0h UT to 13.23 at 4.6h
UT, a similar decrease was also observed in the V band.
The decreasing trend is also clearly evident in the first segment of the soft 
X-ray light curve (Figure 2) and continued in this band for about six hours.

A much better simultaneous coverage was achieved during the 
1998 November 7 observation when the source was followed for about 
12 hours in three optical bands and in the X-rays.
The optical light curves are reported in Figure 3 (upper panel) and show 
a very similar behaviour with a variation of about 0.1 mag over a 
time interval of a few hours: the flux decreased from about 19h UT to a minimum 
at 24h UT and then increased again reaching the previous level in about
two hours.
 
The same general behaviour is clearly apparent in the 1.3-3.0 keV MECS 
data (Figure 3, lower panel), while at higher energies (5-10 keV) 
the count rate, similarly to the 1996 observation, is consistent with 
a constant value. 
The light curves of Figure 3 indicate the possibility that optical and X-ray
variations are not strictly simultaneous: the soft X-ray luminosity could be
declining since the beginning of the observation and could reach
its lowest level about 2-3 hours before the optical minimum.
The following rise, however, seems to start at the same time (about 25h)
in all bands.
This conclusion, however, cannot be firmly established because of
poor statistics.
% and the absence of a lag between the optical and X-ray curves
%within a resolution of about 1.5 hours (1 \sax orbit) cannot be excluded.

Since the soft and hard X-rays follow different evolutions, the X-ray
spectral shape must change with source intensity. 
This is clearly apparent 
from Figure 4 where the MECS hardness ratio (1.5-3.0/3.0-10 keV) in both \sax 
observations is anti-correlated with intensity: the spectrum steepens when 
the source brightens. 
The LECS lightcurve is consistent with these findings, although, in this case 
the photon statistics is poor due to the lower exposure (Table 1) 
and less sensitivity of this instrument compared to the MECS. 

%A visual inspection of Figure 3 suggests that the source could reach 
%its lowest X-ray intensity about 2-3 hours before
%the optical minimum, but this conclusion cannot be firmly established 
%because of poor statistics. The following rise, however, seems to start
%at the same time (about 25h) in all bands.
%In conclusion, no clear lag between the optical and X-ray curves seems 
%to be present within a resolution of about 1.5 hours (1 \sax orbit).

The correlation between the low energy X-ray and optical light curves 
observed by us is in contrast with the results of the 1996 April campaign 
on S5~0716+714 (Otterbein et al. 1998), but agrees with the 1991
observation (Wagner et al. 1996).

Notice that the amplitude of the X-ray variation (a factor of $\approx 2$) 
is higher than at optical-IR frequencies, where it is $\sim$10 \%.

\begin{figure}
\epsfig{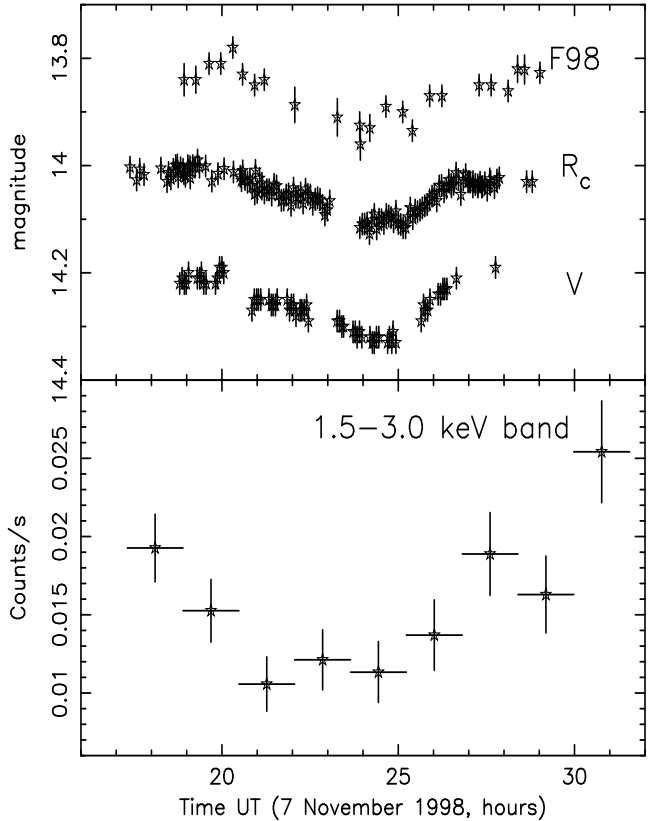}
\caption{Simultaneous optical and 1.5-3.0 keV light curves of S5~0716+714 
during the 1998 observation. A magnitude offset of 0.25 and 0.75 has been
added to the R and F98 data.}
\label{fig3}
\end{figure}

\section{Discussion}

During the two \sax and optical observations presented here S5~0716+714 
was detected in a faint state similar to that seen during other X-ray 
observations when the source was not flaring (Wagner \etal 1996, 
Otterbein \etal 1998). 
Nevertheless significant flux and spectral changes were detected, confirming 
the tendency of this object to be frequently variable. 

The X-ray spectrum of S5~0716+714 can be well represented by a broken power 
law with steep (energy) slope ($\alpha_x\gsim 1.5 $) until  
2-3 keV where a much harder component ($\alpha_x \sim 0.8-0.9 $) emerges.

Figure 5 shows the broad band $\nu-\nu~f_\nu$ SED of S5~0716+714, derived 
from our simultaneous optical and X-ray data together with nearly simultaneous 
radio data from the UMRAO database (Aller \etal 1999) and from 
non-simultaneous photometric data taken from NED. Figure 5 shows that the 
break detected by \sax in the soft X-ray spectrum marks the merging of 
the steep tail of the synchrotron emission into the harder inverse 
Compton component. A similar transition between synchrotron and Compton 
emission has been recently detected in \sax observations of 
another intermediate BL Lac object: ON 231 (Tagliaferri et al. 1999).
The different spectral slopes detected in different 
luminosity states are then easily explained: the spectrum is
steeper when the source is bright and the tail of the synchrotron
component dominates the X-ray flux. When the source is faint 
most of the X-ray flux is due to the flat Compton emission. 
The fast flux variations detected only in the soft X-rays  
strongly suggest that rapid variability 
(correlated with the optical) comes from the high energy tail of the
synchrotron component which peaks at a few times $10^{14} $Hz.  

\begin{figure}
\epsfig{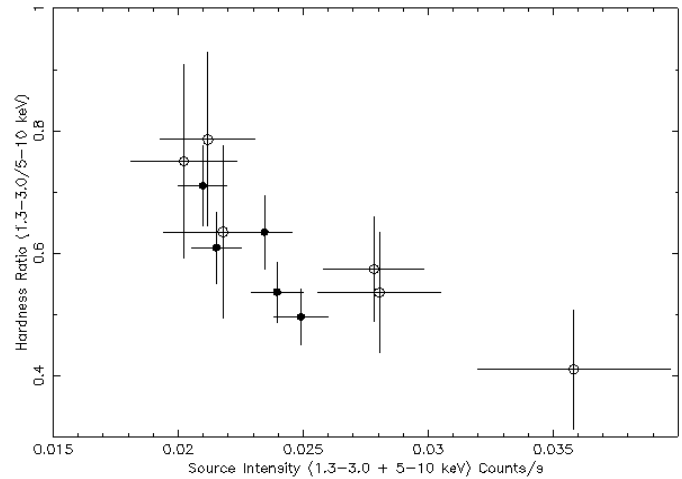}
\caption{The (1.5-3.0/5-10 kev) hardness ratio of S5~0716+714 as a function 
of total intensity in the two bands during the 1996 (filled circles) and 
1998 (open circles) observations.}
\label{fig4}
\end{figure}

Long term variability in the hard Compton component is 
instead apparent from the comparison of the 1996 and 1998 spectral 
energy distributions (Figure 5).

The different variability timescales in the synchrotron (soft X-ray) 
and Compton (harder X-rays) emission explain 
in a natural way the surprisingly poor correlation between the 
simultaneous RXTE (2-10 keV) and ROSAT HRI (0.1-2.4 keV) light 
curves reported by Otterbein et al. (1998).   

Evidence for spectral steepening with intensity was found in the form of  
an anticorrelation between hardness ratio and source flux (Figure 4). 
Such a behaviour is opposite to that observed in many BL Lacs (Giommi 
\etal 1990, Takahashi \etal 1996, 1999) but can be simply explained as due 
to variability in the synchrotron tail that steepens the overall X-ray 
spectrum when the flux of this component increases.
This effect was predicted by Padovani and Giommi (1996) 
to be detectable in intermediate BL Lac objects like S5~0716+714.

\begin{figure*}
%\center{\epsfig{figure=S50716_sed.ps,height=15.0cm,width=10.cm,angle=-90}}
\center{\epsfig{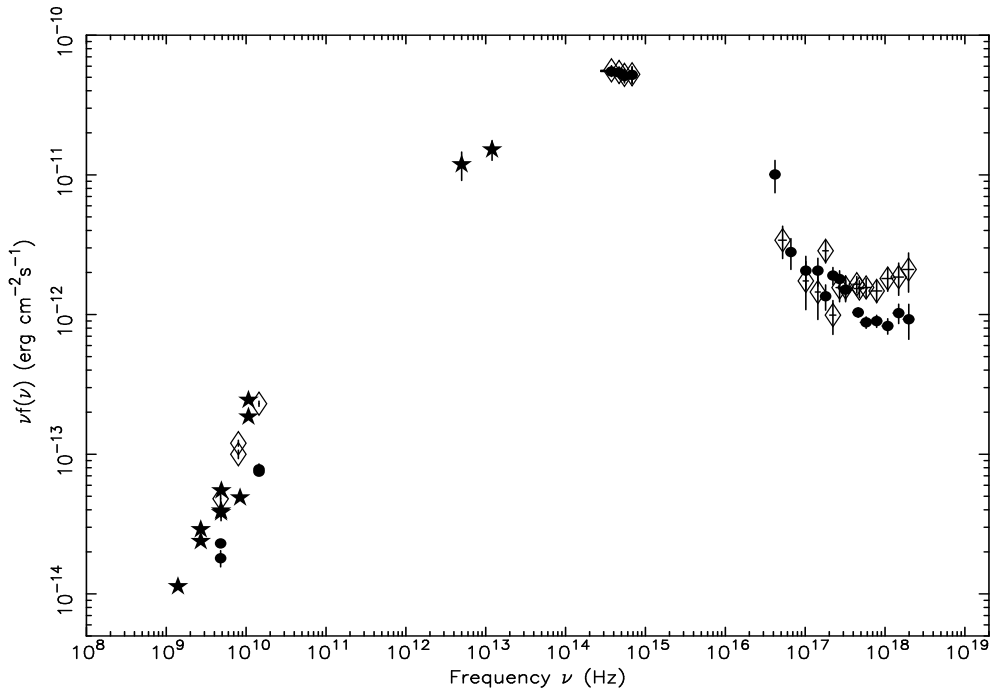}}
\caption{The radio to X-ray Spectral Energy Distribution of S5~0716+714.
The optical and X-ray data are simultaneous and have been collected by us 
in 1996 (filled circles) and in 1998 (open diamonds);
near-simultaneous radio data (filled circles and open diamonds) are from 
the UMRAO database, all other points (filled stars) are from NED. 
On both occasions the \sax data clearly show a break between the tail of 
the synchrotron emission and the on-set of the harder Compton emission 
that is probably responsible for the gamma ray emission.}
\label{fig5}
\end{figure*}

If the optical and X-ray variability are associated, 
the presence of a lag between the minimum of the light curves
in the different bands of the 1998 campaign
can be ascribed to the different radiative cooling
time of the electrons emitting in the optical and in the X-ray bands.
If, instead, the delay is due to geometry (e.g. electrons diffusing out
from an injection region while cooling, or a shock travelling in the jet),
the cooling time must be shorter than the observed lag.
In any case, a firm limit can be derived, by requiring
that the cooling time is equal or shorter than the observed time lag.

We therefore require that the synchrotron and self Compton cooling time
for optical emitting electrons
is equal to or shorter than $t_{lag}\delta/(1+z)$:
\begin{equation}
t_{cool}\, = \, { 6\pi mc^2 \over \sigma_Tc\gamma_o B^2(1+U_S/U_B)}
\, \le { t_{lag}\delta \over (1+z)},
\end{equation}
where $\sigma_T$ is the Thomson scattering cross section, $\gamma_o m_ec^2$
is the energy of the particles emitting at the observed frequency, 
$U_B=B^2/(8\pi)$
is the magnetic energy density, and $U_S$ is the synchrotron radiation 
energy density (as measured in the comoving frame), and $\delta$ is
the beaming or Doppler factor.
The energy $\gamma_0$ is related to the observed frequency by
$\nu_o \simeq 3.7\times 10^6 \gamma_o^2 B \delta/(1+z)$ Hz
(assuming synchrotron radiation).
We then obtain the two limits:
\begin{equation}
B\, \ge  \, 5.6 \, \nu_{15}^{-1/3} t_h^{-2/3}(1+U_S/U_B)^{-2/3}
\left({1+z \over \delta}\right)^{1/3} \quad {\rm G}
\end{equation}
\begin{equation}
\gamma_o\, \le \, 6.9\times 10^3 \nu_{15}^{2/3}
\left[{ t_h (1+U_S/U_B)(1+z) \over \delta }\right]^{1/3} 
\end{equation}
where $t_h$ is the lag time
measured in hours and $\nu_o=10^{15}\nu_{15}$ Hz.
Assuming $\delta=10$, $t_h=4$, $\nu_{15}=0.5$ and $U_S/U_B=1$,
we find $B\ge 0.9$ G and $\gamma_o\le 4.4\times 10^3$, in very good 
agreement with the estimates of Ghisellini \etal (1997).
With these parameters, we expect that the peak of the self
Compton emission is at $h\nu_c\sim \gamma_o^2h\nu_o \lsim 50$ MeV.

It has been shown in a number of occasions that the peak of the 
synchrotron power in BL Lacs tend to move to higher energies during flares 
(e.g. Giommi \etal 1990, Pian \etal 1998, Takahashi \etal 1999, 
Malizia \etal 1999, Giommi \etal 1999, Massaro \etal 1999b).
In the case of S5~0716+714, a shift of $\nu_{peak}$ from the optical band
to a somewhat higher energy, would cause a much larger synchrotron contribution 
to the X-ray flux. 
In case of large events, involving a shift of $\nu_{peak}$ of a factor 10 
or more (as observed in MKN501, Pian et al. 1998, and 1ES2344+514, 
Giommi \etal 1999),  the steep synchrotron emission would entirely 
dominate the X-ray spectrum and further flux increases would cause spectral 
flattening just as observed in many HBL BL Lacs where $\nu_{peak}$ is at
the UV/X-ray frequencies. 

The results presented in this paper demonstrate that intermediate BL Lacs 
are potentially very important to study the relation and the interaction 
between the two main components in the Spectral Energy Distribution of 
BL Lacs. Important observational quantities are the determination of
the position of synchrotron peak at different intensity levels, and the
detailed evolution of the spectral shape in the X-ray band which includes 
in different mixtures both the Synchrotron and Compton components. 
Multi-wavelength campaigns on this type of BL Lacs, covering  
the near infrared and X-rays can lead to a deep understanding of the
physical processes powering Blazars and can unravel the origin of the 
seed photons upscattered to high energies. 

\begin{acknowledgements}
This research has made use of the ASI-BeppoSAX SDC on-line database 
and archive system and of the NASA/IPAC National Extragalactic 
Database (NED).
We thank F. Fiore for providing the code to de-redden the XSPEC unfolded 
spectra used to construct the Spectral Energy Distribution of S5~0716+714.
This research has made use of data from the University of Michigan Radio 
Astronomy Observatory. 
\end{acknowledgements}

\end{document}